\documentclass[12pt]{article}

\usepackage{graphicx}
\usepackage{amsmath}
\usepackage{amsfonts}
\usepackage{amssymb}
\usepackage{bbm}
\usepackage{hyperref}

\newcommand\lsim{\mathrel{\rlap{\lower4pt\hbox{\hskip1pt$\sim$}}
    \raise1pt\hbox{$<$}}}
\newcommand\gsim{\mathrel{\rlap{\lower4pt\hbox{\hskip1pt$\sim$}}
    \raise1pt\hbox{$>$}}}

\newcommand{\beq}{\begin{equation}}
\newcommand{\eeq}{\end{equation}}
\newcommand{\bea}{\begin{eqnarray}}
\newcommand{\eea}{\end{eqnarray}}
\newcommand{\noi}{\noindent}

\newcommand{\TeV}{\,\mathrm{TeV}}
\newcommand{\GeV}{\,\mathrm{GeV}}

\newcommand{\SM}{$SU(3)_C \times \protect 
        \linebreak[0]SU(2)_L \times \protect\linebreak[0]U(1)_Y$}

\begin{document}

\title{\bf  Supersymmetry breaking induced \\ by radiative corrections}
\date{February 13, 2012}
\author{Borut Bajc$^{a,b}$, St\'ephane Lavignac$^{c}$ and Timon Mede$^{a}$
\\
\\ 
$^a$ Jo\v zef Stefan Institute, 1000 Ljubljana, Slovenia\\
$^b$ Department of Physics, University of Ljubljana,\\
1000 Ljubljana, Slovenia\\
$^c$ Institut de Physique Th\'eorique\footnote{Laboratoire
de la Direction des Sciences de la Mati\`ere du Commissariat \`a l'Energie
Atomique et Unit\'e de Recherche associ\'ee au CNRS (URA 2306).} , CEA-Saclay,\\
F-91191 Gif-sur-Yvette Cedex, France}

\maketitle

\centerline{\bf Abstract}

We show that simultaneous gauge and supersymmetry breaking
can be induced by radiative corrections, \`a la Coleman-Weinberg.
When a certain correlation among the superpotential parameters
is present, a local supersymmetry-breaking minimum is found
in the effective potential of a gauge non-singlet field,
in a region where the tree-level potential is almost flat.
Supersymmetry breaking is then transmitted to the MSSM through
gauge and chiral messenger loops, thus avoiding the suppression
of gaugino masses characteristic of direct gauge mediation
models. The use of a single field ensures that no dangerous
tachyonic scalar masses are generated at the one-loop level. 
We illustrate this mechanism with an explicit example based on an
SU(5) model with a single adjoint. An interesting feature of the scenario
is that the GUT scale is increased with respect to standard unification,
thus allowing for a larger colour Higgs triplet mass, as preferred
by the experimental lower bound on the proton lifetime.

%\vspace{0.5cm}
\vfill
\eject

%%%%%%%%%%%
\section{Introduction}
%%%%%%%%%%%

Supersymmetric Grand Unification, although appealing, is plagued by
large uncertainties associated with the necessity of breaking supersymmetry. 
It is an old idea~ \cite{Witten:1981kv} to try and unify the supersymmetry
and the gauge symmetry breaking sectors, thus minimizing and
optimizing them. To our knowledge, apart from Ref.~\cite{Bajc:2008vk},
all attempts so far were using gauge singlets (see for example 
Refs.~\cite{Witten:1981kv,Dine:1982zb,Dimopoulos:1982gm,Banks:1982mg,Derendinger:1982tq,Kaplunovsky:1983yx})
and/or were following nonperturbative approaches (see for example 
Refs.~\cite{Murayama:1997pb,Dimopoulos:1997ww,Luty:1997ny,Agashe:1998kg,Agashe:2000ay}).
In this paper we would like to investigate further the possibility
of breaking simultaneously global supersymmetry and the gauge
symmetry in the framework of perturbative gauge theories without
gauge singlets.

There are several good reasons for using the same field(s) to
spontaneously break supersymmetry and the gauge symmetry.
The first obvious motivation is economy:
there is no need for two separate sectors. The second reason is that this
automatically provides a mechanism for mediating supersymmetry breaking
to the observable sector, with the heavy gauge fields playing the role of
messenger fields. The MSSM soft terms can then be predicted in terms
of a small number of parameters (up to supergravity corrections), and as
a bonus the gaugino mass problem of gauge mediation with chiral
messenger fields\footnote{Namely the fact, already noticed in the
early literature, that
gaugino masses vanish at leading order in direct gauge mediation
models with tree-level supersymmetry breaking~\cite{Komargodski:2009jf}.
Radiative corrections may however solve this
problem~\cite{Giveon:2009yu,Dudas:2010qg}.}
is absent. Finally, another motivation is that the messenger mass scale
is identified with the unification scale, thus helping to avoid a possible
stability problem of the Minkowski vacuum in which supersymmetry
is broken~\cite{Dvali:2011wk,Bajc:2011iu}.

If one sticks to perturbative approaches, several identical representations
(e.g. adjoints\footnote{A single adjoint is not enough
to break both supersymmetry and the unified gauge symmetry at the tree level,
even in the presence of gauge singlets.}) of the unified gauge group are
required if supersymmetry is broken at the tree level, as is usually assumed.
The minimal model of this kind is the SU(5) Grand Unified
example with two adjoints given in Ref.~\cite{Bajc:2008vk}.
The presence of several adjoints, however, generally implies tachyonic
sfermion masses at the
one-loop level~\cite{Intriligator:2010be}. Since gaugino masses arise
at the same order of perturbation theory, this problem cannot be cured
by the renormalization group running between the messenger scale
and low energy. While there are various ways to make this negative
contribution subdominant compared with other
two-loop contributions (for example by imposing a 
mild hierarchy between the vevs of the two fields),
it is difficult to build a consistent model.

In this paper, we avoid this problem by considering the possibility
that supersymmetry and the gauge symmetry are simultaneously
broken by radiative corrections, \`a la
Coleman-Weinberg~\cite{Coleman:1973jx}.
This makes it possible to use a single representation for this purpose,
thus avoiding large negative one-loop
gauge messenger contributions to soft scalar masses\footnote{Note
that in the case of SO(10) gauge symmetry one usually needs more
than one representation to spontaneously break the gauge group,
so that negative one-loop scalar masses are generated and must
be suppressed. }.
Furthermore, we do not make use of gauge singlet fields.

The plan of the paper is the following. In Section~\ref{method}, we
describe the  general method for finding gauge and supersymmetry
breaking minima in the one-loop effective potential of a  single field.
This is then applied to the case of the adjoint representation of SU(5)
in Section~\ref{su5}, where the conditions for the mechanism to work
are identified. In the next sections, we discuss several consequences
of implementing this mechanism in an SU(5) unified theory, taking
as an example the minimal model (to be considered as an existence
proof) in which it can work. 
We first show in Section~\ref{light} that intermediate-mass states are unavoidably
present, which affects the running of the Standard Model gauge couplings
and raises the GUT scale, thus allowing to increase the mass of the colour
Higgs triplet that mediates proton decay. We then comment
on the fermion mass relations and on the doublet-triplet splitting
in Section~\ref{sec:aspects}. Finally, we discuss
the superpartner spectrum in Section~\ref{sec:spectrum}
and give our conclusions in Section~\ref{conclusions}.

%%%%%%%%%%%%%%%%%%%
\section{\label{method}The method}
%%%%%%%%%%%%%%%%%%%

The method we will use to seek local supersymmetry breaking minima
in the one-loop effective potential can be seen as a generalized
version of Witten's mechanism for generating
the hierarchy between the weak scale and the GUT scale~\cite{Witten:1981kv}.
Witten's method  consists in finding a flat direction that breaks both
supersymmetry and the gauge group, and in stabilizing it 
by radiative corrections. At the one-loop level the effective 
potential has the following approximate form:
\beq
V(\sigma)=\frac{\left|F\right|^2}{1+
\left(c_g-c_\lambda\right)
\log{(|\sigma|^2/\mu^2)}}\ ,
\label{eq:V_Witten}
\eeq
where $c_x=c_x^0 x^2/16\pi^2$ ($x=g$ or $\lambda$) 
are loop-suppressed factors, $\sigma$ 
is the flat direction, and $\mu$ is the renormalization scale.
For the mechanism to work, it is crucial that both
$g$ and $\lambda$ participate. Typically, the gauge coupling
goes at high energy towards asymptotic freedom while the
Yukawa coupling behaves the opposite way, so that
$c_g-c_\lambda$ changes sign and becomes negative above
some scale, thus stabilizing the value of $\sigma$ (which otherwise
would be pushed to higher values by $c_g-c_\lambda > 0$).
The minimum is reached at $|\sigma|=\mu_0$ where 
\beq
\label{min}
c_g(\mu_0)=c_\lambda(\mu_0)\, .
\eeq
Notice that $F$ in Eq.~(\ref{eq:V_Witten}) is a constant
($\sigma$ is a tree-level flat direction). Witten's mechanism
cannot work if there is no flat direction (this is the typical situation 
in the minimal case of a single field\footnote{An exception to this rule is 
the cubic superpotential of the adjoint of SU(6).}), 
or when for some reason the minimum (\ref{min}) cannot be reached.

The method we are going to use in this paper relies on the
observation that a minimum at a finite value of $\sigma$ can also
appear in the case where $F$ is a slowly varying function of $\sigma$.
As we will see, this can work even for negligible 
Yukawa couplings, i.e. for an effective potential of the form:
\beq
V(\sigma)=\frac{\left|F(\sigma)\right|^2}{1+c_g 
\log{(|\sigma|^2/\mu^2)}}\ .
\eeq
In the first approximation one could even neglect the scale dependence
of $c_g$, i.e. of  the  gauge coupling, since its change is not crucial
anymore. It is now the field dependence in $F(\sigma)$ that counterbalances
the one-loop logarithmic contribution 
in the denominator. For this to work, the tree-level potential should be almost flat. 
In practice this is the case when the first few derivatives of the superpotential are 
small. We are not even close to an extremum of the tree-level potential,
but this is corrected by the one-loop logarithmic contribution. 
In supersymmetric theories one often expands the scalar potential
around tree-level flat directions, but this case is actually closer to the
original Coleman-Weinberg philosophy, in which one expands the potential
in a region where the first tree-level derivatives are non-vanishing.

%%%%%%%%%%%%%%%%%%%%%%%%%%%%%%%%
\section{\label{su5} An SU(5) example with a single adjoint}
%%%%%%%%%%%%%%%%%%%%%%%%%%%%%%%%

Let us illustrate the method exposed in the previous section with
an SU(5) example. The chiral superfield $\sigma$ is identified
with the Standard Model singlet direction of an adjoint $\Sigma$:
\beq
\Sigma=\frac{\sigma}{\sqrt{30}}\, {\rm diag} (2,2,2,-3,-3) + \cdots\, .
\eeq
In the small $|F(\sigma)/\sigma^2|$ limit, which is the phenomenologically
relevant one\footnote{Since the messenger fields are the components
of the heavy gauge supermultiplets, their mass scale is
$v \equiv \langle \sigma \rangle \sim M_{GUT}$ and the superpartner masses
are given by $m_{\rm soft} \sim (g^2/16\pi^2) F(v)/v$. Hence $F(v)$
must be much smaller than $v^2$ if supersymmetry is realized
at low energy.}, the one-loop effective potential for $\sigma$ is given
to a very good approximation by:
\beq
V(\sigma)=\frac{|F(\sigma)|^2}{1+c\log{(|\sigma|^2/\mu^2)}}\ ,
\label{eq:V_eff}
\eeq
were $c$ only receives gauge contributions and is therefore positive.
We are looking for an SU(5)-breaking local minimum
$\langle\sigma\rangle = v \neq 0$
in which supersymmetry is broken, i.e. $F(v) \neq 0$.
The extremum condition reads:
\beq
\label{f1}
F'(v)=cF(v)/v\, ,
\eeq
where the renormalization scale has been chosen to be $\mu = v$,
while the requirement that the extremum is a minimum tells us that
\beq
  \left|\, \frac{c^2}{v^2}\, F(v) \right|\ \geq\
  \left|\, F'' + \frac{c}{v^2}\, F(v) \right| .
\eeq
The general solution to this constraint can be parametrized as: 
\beq
\label{f2}
F''(v)=(ac^2-c)F(v)/v^2\, ,  \qquad  |a| \leq 1\, .
\eeq

The conditions~(\ref{f1}) and~(\ref{f2}) require some level of
correlation between the Lagrangian parameters.
To make this correlation explicit, we can expand:
\beq
F(\sigma)=\sum_{n=0}^{\infty}\, \frac{F^{(n)}(v)}{n!}\, (\sigma-v)^n\, ,
\label{eq:Taylor}
\eeq
and use Eqs.~(\ref{f1}) and~(\ref{f2}) to express $F'(v)$
and $F''(v)$ on the right-hand side in terms of $F(v)$ and $v$.
Let us see this on the following example (the presence of non-renormalizable
terms in the superpotential will be justified later):
\beq
W = \frac{\mu}{2}\, \mbox{Tr}\, \Sigma^2
  + \sqrt{30}\, \frac{\lambda}{3}\, \mbox{Tr}\, \Sigma^3
  + \frac{\kappa_1}{4M}\, \mbox{Tr}\, \Sigma^4
  + \frac{\kappa_2}{4M} \left( \mbox{Tr}\, \Sigma^2 \right)^2 .
\label{eq:W_Sigma}
\eeq
We thus have:
\beq
F(\sigma)=\mu\sigma-\lambda\sigma^2+\frac{\kappa}{M}\sigma^3\, ,
\label{eq:Fsigma}
\eeq
where
\beq
\kappa = \frac{7}{30}\, \kappa_1 + \kappa_2\, .
\eeq
\begin{figure}[t]
\begin{center}
\includegraphics[width=10cm]{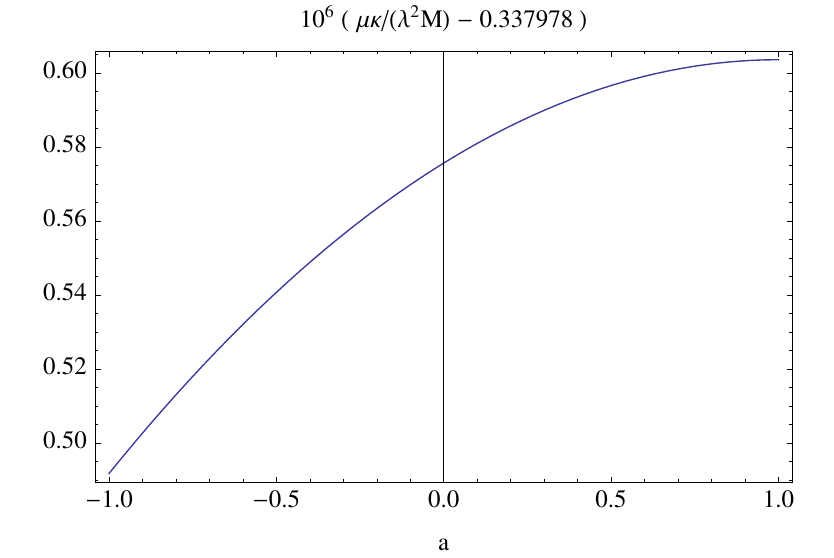}
\caption{\label{fig1}The allowed values of the combination 
$\mu\kappa/(\lambda^2 M)$ for $c=0.04$ show the amount of fine-tuning
needed among superpotential parameters in order to obtain a metastable
supersymmetry breaking minimum with a single adjoint of SU(5).}
\end{center}
\end{figure}
Inserting Eqs.~(\ref{f1}), (\ref{f2}) and~(\ref{eq:Fsigma}) into
Eq.~(\ref{eq:Taylor}), we obtain:
\bea
\mu&=&\frac{F(v)}{2v}\left(6-5c+ac^2\right) ,  \label{eq:mu}  \\
\lambda&=&\frac{F(v)}{v^2}\left(3-4c+ac^2\right) ,  \label{eq:lambda}  \\
\frac{\kappa}{M}&=&\frac{F(v)}{2v^3}\left(2-3c+ac^2\right) . \label{eq:kappa}
\eea
Since $F(v) \ll v^2$, all superpotential parameters must be  suppressed:
$\mu \ll M$, $\lambda \ll 1$, $\kappa\, v/M \ll 1$. We can therefore calculate $c$
neglecting all superpotential parameters, in which case only the $X$ and $Y$
massive vector supermultiplets contribute to the effective potential, with masses
$m_X = m_Y = \frac{5}{6}\, g^2_5 |\sigma|^2$, where $g_5$ is the $SU(5)$ gauge
coupling. In the approximation of small $|F(\sigma) / \sigma^2|$, this yields
$V(\sigma) = |F(\sigma)|^2 \left( 1 + \frac{12}{16\pi^2}\, \frac{5}{6}\, g^2_5
\left[ \log (\frac{5}{6}\, g^2_5 |\sigma|^2 / \Lambda^2) + 1 \right] \right)^{-1}$
(where $12$ is the multiplicity of the $X$ and $Y$ gauge fields),
which to a very good approximation leads to Eq.~(\ref{eq:V_eff}) with
\beq
c\, =\, \frac{12}{16 \pi^2}\, \frac{5}{6}\, g^2_5\, =\, 10\, \frac{\alpha_{GUT}}{4\pi}\, \approx\, 0.04\, ,
\eeq
where $\alpha_{GUT} = g^2_5/(4\pi)$.
The correlation between $\mu$, $\lambda$ and $\kappa/M$
arises from the fact that $c$ is a small loop-suppressed parameter,
and that $a$ is bounded to have a modulus
less or equal to $1$.
This is shown in Fig.~\ref{fig1}, where the
combination $\mu\kappa/(\lambda^2 M)$ is plotted as a function
of $a$ for $c=0.04$. 
The corresponding fine-tuning can be estimated by 
$d \log{(\mu\kappa/(\lambda^2 M))}/d a|_{a=0}\simeq c^4/18 = {\cal O}(10^{-7})$.

\begin{figure}[t]
\begin{center}
\hspace{-1.cm}
\includegraphics[width=8.cm]{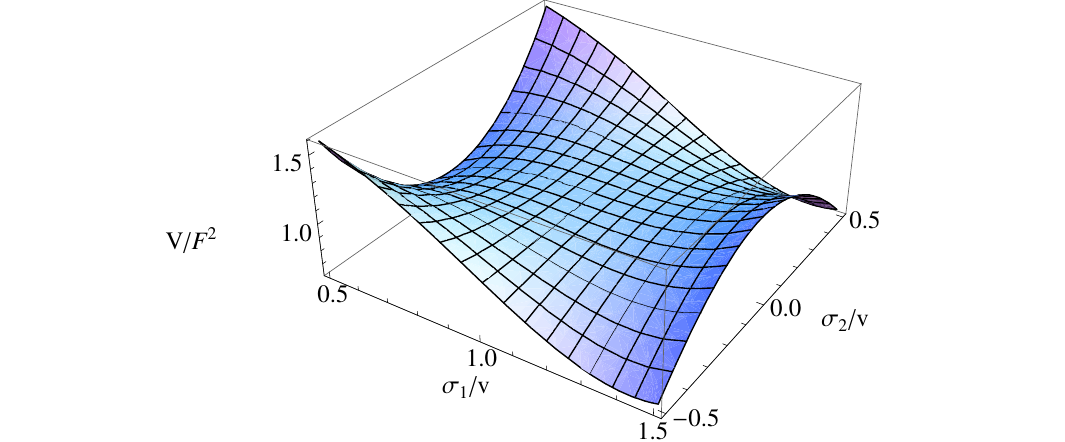}
\hspace{-2.cm}
\includegraphics[width=8.cm]{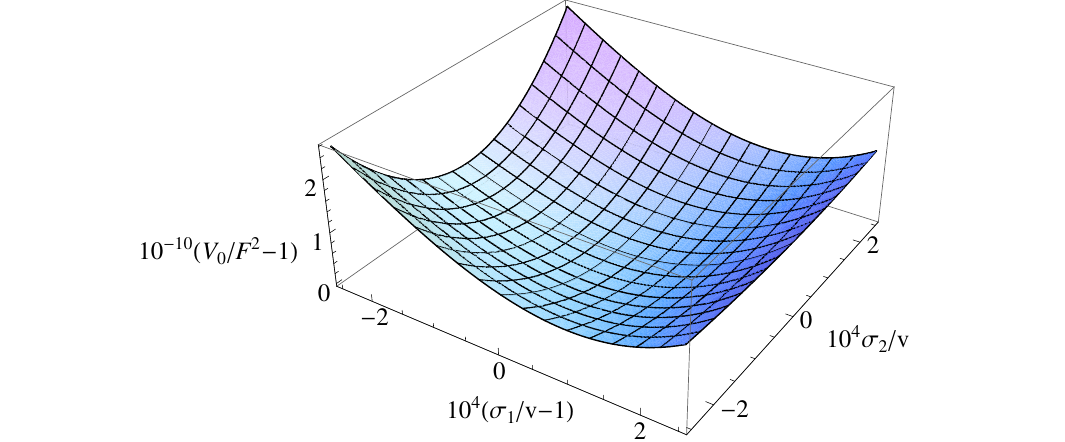}
\caption{\label{fig2} The one-loop effective potential for $\sigma$
in a larger (left) and smaller (right) region around the local minimum,
for $a=0$ and $c=0.04$. $\sigma_1$ and $\sigma_2$ are the real
and imaginary parts of $\sigma$, respectively.}
\end{center}
\end{figure}

Fig.~\ref{fig2} shows the appearance of a local supersymmetry
breaking minimum in the one-loop effective potential when
conditions~(\ref{f1}) and~(\ref{f2}), or equivalently
Eqs.~(\ref{eq:mu})--(\ref{eq:kappa}), are satisfied.
If we restore the tree-level value $c=0$, the left figure remains 
visibly identical, while the right one displays a saddle point
instead of a local minimum, as it should.

One may wonder whether the non-renormalizable terms in
the superpotential~(\ref{eq:W_Sigma}) are really necessary
for our purpose. The renormalizable case would imply $\kappa = 0$,
or due to Eq.~(\ref{eq:kappa}):
\beq
ac^2-3c+2=0\quad \Rightarrow\quad c>0.56\, ,
\eeq
hence the model would become non-perturbative (remember that
$c$ contains a loop factor). Thus perturbativity forces us to a superpotential
with non-renormalizable operators.

%%%%%%%%%%%%%%%%%%%%%%%%%
\section{\label{light} Intermediate states, the GUT scale and proton decay}
%%%%%%%%%%%%%%%%%%%%%%%%%
 
The fact that the superpotential parameters $\mu$, $\lambda$
and $\kappa$ must be suppressed to ensure $F(v) \ll v^2$
has important consequences for the mass spectrum and for gauge
coupling unification, similarly to the case of tree-level supersymmetry
breaking in a renormalizable model with two adjoints~\cite{Bajc:2008vk}
and in the supergravity model with a single adjoint of Ref.~\cite{Bajc:2006pa}.
Indeed, $\Sigma$ contains a weak triplet $\Sigma_3$ and a colour
octet $\Sigma_8$ whose masses $m_3$ and $m_8$ are determined
by the parameters $\mu$, $\lambda$, $\kappa_1$ and $\kappa_2$.
Since, according to Eqs.~(\ref{eq:mu})--(\ref{eq:lambda}),
\beq
\mu\, \sim\, \lambda v\, \sim\, \frac{F(v)}{v} \sim \left( \frac{\alpha}{4\pi} \right)^{\! -1}\!
m_{\rm soft}\, ,
\eeq
only $\kappa_1$ and $\kappa_2$ can give a potentially large
(i.e. much larger than the superpartner mass scale $m_{\rm soft}$)
contribution to $m_3$ and $m_8$. Indeed, while the combination
$\kappa = 7 \kappa_1 / 30 + \kappa_2$ is constrained by
Eq.~(\ref{eq:kappa}) to be small, $\kappa v^2/M \sim F(v)/v$,
$\kappa_1$ and $\kappa_2$ could in principle be much larger.
Even in the case of a strong cancellation
between $\kappa_1$ and $\kappa_2$, however, $m_3$ and
$m_8$ are at most of order $\kappa_{1,2} v^2 / M \lesssim M^2_{GUT}/M$,
i.e. below the GUT scale, which affects gauge coupling unification.

More precisely, in the absence of a cancellation between
$\kappa_1$ and $\kappa_2$, the colour octet and weak triplet are light
with masses of order $F(v)/v$. Gauge couplings can still unify with
superpartner masses in the TeV range, but this happens above
the Planck scale. We are thus led to consider the fine-tuned case in which
$\kappa_1 \simeq -30 \kappa_2 / 7 \gg \kappa$ in order to allow
for unification below the Planck scale. The triplet and octet masses
are then given by (identifying $M$ with the Planck mass $M_P$):
\beq
m_3\, \simeq\, 4 m_8\, \simeq\, \frac{2}{3}\, \kappa_1 \frac{v^2}{M}\, .
\eeq
Inspection of the renormalization group equations (RGEs) shows that
unification is achieved for the following values of the GUT scale and
colour Higgs triplet\footnote{We are referring here to the SU(5) partners
of the MSSM Higgs doublets, which should not be confused with
the weak triplet $\Sigma_3$ living in the adjoint representation.}
mass~\cite{Bachas:1995yt,Chkareuli:1998wi,Bajc:2002pg}:
\beq
M_{GUT}\, =\, M_{GUT}^0\left(\frac{M_{GUT}^0}{\sqrt{m_3m_8}}\right)^{\! 1/2}\!
  =\, M_{GUT}^0\left(\frac{3M_P}{\kappa_1 M_{GUT}^0}\right)^{\! 1/4} ,
  \label{eq:MGUT}
\eeq
\beq
m_T\, =\, m_T^0\left(\frac{m_3}{m_8}\right)^{\! 5/2}\! =\, 32 m^0_T\, ,
\label{eq:mT}
\eeq
where $M^0_{GUT}$ and $m^0_T$ refer to the values of the GUT scale
and Higgs triplet mass in the case $m_3 = m_8$. 
The RGE analysis of Ref.~\cite{Murayama:2001ur}
gives\footnote{Strictly speaking, this number is obtained from a partial
2-loop analysis, and should not be used in our 1-loop approximation. 
A more accurate analysis would give slightly different values
for $M_{GUT}$ and $m_T$ without changing the qualitative picture.}
$m_T^0 \leq 3.6\times 10^{15} \GeV$ ($90\%\, \mbox{C.L.}$), while
$M_{GUT}^0 \approx 2\times 10^{16} \GeV$. Hence the triplet mass
can be increased to $10^{17} \GeV$, a welcome feature for proton
decay. Using Eq.~(\ref{eq:MGUT}), we can see that
$M_{GUT} \approx m_T \approx 10^{17} \GeV$ is obtained for
$\kappa_1 \approx 0.3$.
An important feature of the model is the increase of the GUT scale. 
This can be understood by the unavoidable presence of
non-renormalizable terms in the adjoint superpotential. They 
automatically lead to intermediate states which postpone unification, 
as can be seen from Eq.~(\ref{eq:MGUT}).

The amount of fine-tuning between $\kappa_1$ and $\kappa_2$
that is needed in order to allow for large enough intermediate state masses,
so that unification occurs at least one order of magnitude below the Planck
mass, can be estimated by:
\beq
  \frac{\kappa}{\kappa_1}\, \sim\, \frac{F(v) M}{\kappa_1 v^3}\ \lesssim\,
  \left( \frac{\alpha}{4\pi} \right)^{\! -1} \frac{m_{\rm soft} M_P}{\kappa_1 M^2_{GUT}}\,
  \sim\, 10^{-10} \left( \frac{m_{\rm soft}}{1 \TeV} \right) ,
  \label{kappagamma}
\eeq
in which we have inserted $\kappa_1 = 0.3$ and $M_{GUT} = 10^{17} \GeV$.
The inequality in Eq.~(\ref{kappagamma}) takes into account the fact
that soft terms receive other contributions than the gauge-mediated
ones, e.g. from supergravity, and that these may dominate (see
Section~\ref{sec:spectrum}). However we have seen in Section~\ref{su5}
that the parameter $\kappa$ must be correlated with $\mu$ and $\lambda$
with a precision of $10^{-7}$ in order for a supersymmetry breaking
minimum to be induced in the one-loop effective potential of $\sigma$.
Hence one may view these two tunings -- the one needed to ensure
the existence of the supersymmetry breaking minimum and the one
needed to allow for unification below the Planck scale -- as a
single fine-tuning of order $10^{-17}$ between $\kappa_1$ and $\kappa_2$.

The increase of the GUT scale and of the colour triplet mass has several
important consequences. On the one side, a heavier Higgs triplet
implies a stronger suppression of the $D=5$ proton decay operators
and reduces the tension with the experimental
upper bound on the proton lifetime. 
On the other side, $M_{GUT} \sim 10^{17} \GeV$ means that
the supergravity contribution to soft masses will dominate over
the gauge-mediated ones (see the discussion in Section~\ref{sec:spectrum}),
unless one relies on a mechanism like no scale supergravity or
conformal sequestering. Therefore, the predictive power of gauge
mediation for the superpartner spectrum is generally lost. It should
be said though that the situation is exactly similar to the one
of the so-called anomaly-mediated supersymmetry breaking
scenario~\cite{Randall:1998uk,Giudice:1998xp}.
Finally, a larger GUT scale implies a stronger sensitivity
to higher-dimensional operators suppressed by the Planck mass.

Note that one arrives at very similar conclusions in a supergravity model
with a single SU(5) adjoint and a canonical K\"ahler potential~\cite{Bajc:2006pa},
or in a global supersymmetric SU(5) model with two adjoints \cite{Bajc:2008vk}.

Some of these conclusions may be evaded for more complex 
superpotentials\footnote{Another possibility is 
to replace the 24-dimensional adjoint with a 75-dimensional representation.}.
In fact the quartic superpotential of Eq.~(\ref{eq:W_Sigma}) is only
a particular example. The most general superpotential for $\Sigma$
involves a sum of non-renormalizable terms of the form
$\Sigma^n$, $n \geq 4$, i.e. gauge invariants
${\cal O}_{n,k} \equiv \prod_i \left( \mbox{Tr}\, \Sigma^{n_{i,k}} \right)^{a_{i,k}}$ 
with $\sum_i a_{i,k} n_{i,k} = n$:
\beq
W = \frac{\mu}{2}\, \mbox{Tr}\, \Sigma^2
  + \sqrt{30}\, \frac{\lambda}{3}\, \mbox{Tr}\, \Sigma^3
  +\, \sum_{n \geq 4}\, \sum_k\, \kappa_{n,k}\, {\cal O}_{n,k}\, .
\eeq
The constraints discussed in this and the previous sections can still be
satisfied with a single fine-tuning between the coefficients $\kappa_{n,k}$
of the operators corresponding to a single $n$, provided that
the coefficients of the other operators are small.
The tuning needed is the one  that makes (with the help of the small
couplings $\mu$, $\lambda$ and $\kappa_{p,k}$, $p \neq n$) 
the first  three derivatives of the superpotential small enough,
while keeping $m_3$ and $m_8$ large
enough to allow for unification below the Planck scale.
The simplest and most predictive cases are $n=4$ (the case
discussed so far) and $n=5$,
for which there are only 2 independent operators ${\cal O}_{n,k}$,
namely ${\cal O}_{4,1} = \mbox{Tr}\, \Sigma^4$,
${\cal O}_{4,2} = \left(\mbox{Tr}\, \Sigma^2 \right)^2$, and
${\cal O}_{5,1} = \mbox{Tr}\, \Sigma^5$,
${\cal O}_{5,2} = \left(\mbox{Tr}\, \Sigma^3 \right) \left(\mbox{Tr}\, \Sigma^2 \right)$.
The triplet mass is predicted only in these two cases
(with a ratio $m_3/m_8 = 28/3$ for $n=5$).

At this point we have all the necessary ingredients to attack another issue.  
The minimum obtained in the previous section is not the global minimum of the 
effective potential. In the limit of infinite Planck mass, one has to check
whether the lifetime of this metastable vacuum can be
longer than  the age of the Universe.
The transition rate to the SU(5) and supersymmetry preserving
minimum at the origin can be estimated in the usual way by $e^{-S}$, where
$S\approx\Delta\Phi/\Delta V\approx M_{GUT}^4/F^2\approx 10^{20}$.
Hence the metastability of the supersymmetry-breaking minimum
is not an issue.

One could be worried that there are other supersymmetry-breaking or preserving
minima, for example in the SU(4)$\times$U(1) direction. This seems
unlikely because the fine-tuning~(\ref{kappagamma})
between $\kappa_1$ and $\kappa_2$
(which is required in order to satisfy Eq.~(\ref{eq:kappa}) while preventing
$\Sigma_3$ and $\Sigma_8$ to be light) is done in the Standard Model
direction.
In the SU(4)$\times$U(1) direction, the relevant combination
of $\kappa_1$ and $\kappa_2$ is different and gives an effective quartic
coupling of order 1, while $\mu$ and $\lambda$
remain small. This means that the SU(4)$\times$U(1)-invariant
minimum must be very close
to the origin, hence the probability for the Standard Model vacuum to decay to it
is as suppressed as the probability to decay to the global SU(5)-preserving minimum.

Reintroducing a finite Planck scale one would prefer to satisfy
the Coleman-De Luccia bound~\cite{Coleman:1980aw} and thus avoid
a possible instability of the Minkowski vacuum~\cite{Dvali:2011wk}.
Although an exact calculation is beyond the scope of this paper, the vicinity
of the GUT scale to the Planck scale makes this possible, as pointed out
in Ref.~\cite{Bajc:2011iu}.

%%%%%%%%%%%%%%%%%
\section{Model-building aspects}
\label{sec:aspects}
%%%%%%%%%%%%%%%%%

In this section, we address a few model-building issues related
to the quark-lepton mass relations and to the doublet-triplet splitting
problem.

%%%%%%%%%%%%%%%%%%%%%
\subsection{\label{mass}Fermion masses}
%%%%%%%%%%%%%%%%%%%%%

A generic problem of SU(5) models with a minimal Higgs sector
is the equality, at the renormalizable level, between the down quark
and charged lepton Yukawa couplings. As is well known, this
leads to wrong predictions for the $m_\mu/m_s$ and $m_e/m_d$
mass ratios, and the GUT-scale relations must therefore be corrected
for the first two generations (while the small discrepancy between
the observed and predicted $m_\tau / m_b$ ratio
may be obtained from supersymmetric threshold corrections alone).

To do this without introducing new degrees of freedom, one can just
employ higher-dimensional operators obtained by inserting powers
of the adjoint $\Sigma$ in the standard renormalizable Yukawa coupling:
\beq
  W_{Y_e,Y_d}\, =\, (Y_3)_{ij}\, 10_F^i\bar 5_F^j\bar 5_H\, +\,
  (Y_4)_{ij}\, 10_F^i \left(\! \frac{\Sigma}{M_P}\, \bar 5_F^j\! \right) \bar 5_H\, +\, \cdots\, .
\label{eq:W_Yuk}
\eeq
The non-renormalizable operators contribute to both the down-type
Yukawa couplings and A-terms\footnote{They also contribute to the
colour triplet couplings to light fermions, which may help suppress
the proton decay rate.}. For instance, Eq.~(\ref{eq:W_Yuk}) gives:
\beq
\label{deltaW}
  (Y_d)_{ij} = (Y_3)_{ij} + \frac{2}{\sqrt{30}} \frac{v}{M_P}(Y_4)_{ij}\, , \quad
  (Y^T_e)_{ij} = (Y_3)_{ij} - \frac{3}{\sqrt{30}} \frac{v}{M_P}(Y_4)_{ij}\, ,
\eeq
\beq
\label{deltaV}
  (\delta A_d)_{ij}\, =\, \frac{2}{\sqrt{30}} \frac{F(v)}{M_P}(Y_4)_{ij}\, , \qquad
  (\delta A^T_e)_{ij}\, =\, - \frac{3}{\sqrt{30}} \frac{F(v)}{M_P}(Y_4)_{ij}\, ,
\eeq
where the A-terms are defined by $V_{\rm soft} \ni (A_u)_{ij} \tilde Q^i \tilde U^{cj} H_u
+ (A_d)_{ij} \tilde Q^i \tilde D^{cj} H_d$
$+\, (A_e)_{ij} \tilde L^i \tilde E^{cj} H_d + \mbox{h.c.}$
and also receive a contribution from gauge messenger fields
(see Section~\ref{sec:spectrum}). This typically gives large A-terms
for the first and second generations, for which significant corrections
to the renormalizable Yukawa couplings are required:
\beq
  (\delta A_{d,e})_{ij}\, =\, \frac{F(v)}{v}\, (\delta Y_{d,e})_{ij}\ \lesssim\,
   \left( \frac{\alpha}{4\pi} \right)^{\! -1} m_{\rm soft}\, (\delta Y_{d,e})_{ij}\, .
\label{eq:A_size}
\eeq
For the second generation of leptons, one typically needs
$(\delta Y_e)_{22} \sim (Y_e)_{22}$, which yields
$|(\delta A_e)_{22}| \lesssim (Y_e)_{22} (\alpha/4\pi)^{\! -1}\, m_{\rm soft}$,
to be compared with the standard tree-level constraint
$|A_f| \lesssim 3 y_f m_{\rm soft}$ ensuring the absence of deeper
charge and colour beaking (CCB) minima~\cite{Frere:1983ag}. 
Although the upper bound in Eq.~(\ref{eq:A_size}) has little chance
to be saturated, because
gravitational contributions to soft terms are likely to dominate over
the gauge-mediated ones (see Section~\ref{sec:spectrum}), 
such large A-terms are nevertheless at odds with the CCB constraints.
Barring the arguments of Ref.~\cite{Dvali:2011wk},
however, deeper CCB minima are acceptable if the MSSM
vacuum is metastable with respect to these minima. The corresponding
constraints on A-terms are weaker than the ones associated with
exact stability of the MSSM vacuum with respect to CCB minima,
and moreover, they are only relevant for the A-terms associated
with large Yukawa couplings~\cite{Claudson:1983et,Kusenko:1996jn}.

There is actually another type of contribution to fermion masses,
which is particularly important in the case of large A-terms, coming
from low-energy supersymmetric threshold corrections. These alone,
according to Ref.~\cite{Enkhbat:2009jt}, could correct the wrong
SU(5) mass relations. A complete analysis including these threshold
corrections, with the A-terms given by Eq.~(\ref{deltaV}), on top of
the contribution from Eq.~(\ref{deltaW}) is beyond the scope of
this paper.

However, there is more freedom than stated above. Since the GUT
scale is typically large, e.g. for the model of Section~\ref{su5}
$M_{GUT}/M_P\gtrsim 0.05$, non-renormalizable operators
suppressed by $1/M^2_P$ may be relevant even to the determination
of the second generation Yukawa couplings. Schematically, one has:
\beq
  Y\, =\, Y_3 + x \left( c_4 Y_4 + c_5 x Y_5 \right) ,
\eeq
\beq
  \delta A\, =\, \frac{F(v)}{v}\, x \left( c_4 Y_4 + 2 c_5 x Y_5 \right) ,
\label{eq:Aterm_Y5}
\eeq
where $x \equiv M_{GUT}/ \sqrt{30}\, M_P$ and $c_4$, $c_5$ are
coefficients of order 1. Due to the factor of 2 in Eq.~(\ref{eq:Aterm_Y5}),
the correlation between the A-terms and the corrections to the
renormalizable Yukawa couplings is spoiled.
This makes it possible to ensure the stability of the MSSM vacuum
with respect to CCB minima, as required by the arguments of
Refs.~\cite{Dvali:2011wk,Bajc:2011iu}.

%%%%%%%%%%%%%%%%%%%%%%%%
\subsection{\label{dt}Doublet-triplet splitting}
%%%%%%%%%%%%%%%%%%%%%%%%

Since the adjoint $\Sigma$ acquires a VEV along both its lowest and F-term
components, two fine-tunings are a priori necessary: one to split the $\mu$
term from the triplet mass, another one to split the $B\mu$ term
from the scalar triplet-antitriplet mixing parameter, which is of
order $F$. To perform this double splitting, one needs to go beyond
the renormalizable level:
\beq
W\, =\, \bar H \left( M_H + \eta \Sigma+ \frac{\beta_1}{M}\, \Sigma^2
+ \frac{\beta_2}{M}\, \mbox{Tr}\, (\Sigma^2) \right) H\, ,
\eeq
where $H \equiv 5_H$ and $\bar H \equiv \bar 5_H$.
The two fine-tunings are:
\bea
\label{mu}
M_H - \sqrt{\frac{3}{10}}\, \eta\, v +\! \left( \frac{3}{10}\, \beta_1 + \beta_2 \right)\!
  \frac{v^2}{M}\, =\, 0\, ,  \\
\label{bmu}
-\, \sqrt{\frac{3}{10}}\, \eta + 2 \left( \frac{3}{10}\, \beta_1 + \beta_2 \right)\!
  \frac{v}{M}\, =\, 0\, ,
\eea
and as mentioned above require that at least one of the two
parameters $\beta_1$ and $\beta_2$ be non vanishing.
The triplet mass is thus given by:
\bea
\label{mt}
m_T & = & M_H + \frac{2}{\sqrt{30}}\, \eta\, v
  +\! \left( \frac{4}{30}\, \beta_1 + \beta_2 \right)\! \frac{v^2}{M}\,
  \nonumber  \\
& = & \left(\frac{5}{6}\, \beta_1 + \frac{10}{3}\, \beta_2 \right)\! \frac{v^2}{M}\, .
\eea
The doublet-triplet splitting can be performed in two ways: 
\begin{enumerate}
\item 
the first one sticks to the principle of minimal fine-tuning. If one chooses
$M_H=\eta=0$, a single fine-tuning is needed in order to satisfy
both Eqs.~(\ref{mu}) and~(\ref{bmu}), namely $3\beta_1/10 = - \beta_2$.
The triplet mass then reads $m_T=-(\beta_1/6)(M^2_{GUT}/M_P)$.
To achieve $m_T = 10^{17} \GeV$ as in the model of Section~\ref{su5},
one needs $M_{GUT} \simeq 10^{18} \GeV/\sqrt{|\beta_1|}$, so that
the perturbative expansion in powers of $M_{GUT}/M_P$ can no longer
be trusted (either because $M_{GUT}/M_P$ is not a small parameter,
or because $\beta_1$ is non-perturbative). Perturbativity thus
forces one to consider a lower value of the triplet mass, however still
consistent with the RGE analysis of Ref.~\cite{Murayama:2001ur}.
One then has to invoke another way of suppressing proton decay
(see e.g. Ref.~\cite{Bajc:2002pg}).
\item
the second possibility is to choose $\beta_1$ and $\beta_2$ so as
to maximize the triplet mass. This is done at the expense of a second
fine-tuning, the combination $3\beta_1/10 + \beta_2$ being fixed
in Eqs.~(\ref{mu}) and~(\ref{bmu}). For instance,  $m_T = 10^{17} \GeV$
can be achieved with $\beta_1 = \beta_2 = 1$ and
$M_{GUT} = 2.4 \times 10^{17} \GeV$.
\end{enumerate}

An alternative possibility is to use the missing partner
mechanism~\cite{Masiero:1982fe,Grinstein:1982um}. This requires
replacing the adjoint $\Sigma$ with a $75_H$ and adding a pair of
$50_H \oplus \overline{50}_H$. In this case the doublet-triplet
splitting is realized for both $\mu$ and $B\mu$ without fine-tuning.
The spontaneous breaking of supersymmetry and of the SU(5)
gauge symmetry by the $75_H$ proceeds as
described in Section~\ref{su5}, but with a different
value of $c$. The wrong SU(5) mass relations can be corrected
by non-renormalizable operators involving powers of the $75_H$.
The main differences with the adjoint case are
the spectrum of intermediate states left over from the $75_H$,
the running of gauge couplings (Section~\ref{light}), and the
predictions for the MSSM soft terms (Section~\ref{sec:spectrum}).

%%%%%%%%%%%%%%%%%%%
\section{The superpartner spectrum}
\label{sec:spectrum}
%%%%%%%%%%%%%%%%%%%

In this section, we discuss the gauge-mediated contributions to the
MSSM soft terms in the SU(5) model described in Section~\ref{su5}.
Before doing so, let us stress that the
supergravity contributions will generally dominate due
to the large messenger scale, $M_{GUT} \sim 10^{17} \GeV$.
Indeed, the typical size of a gravity-mediated soft term is
$m_{3/2} = F / \sqrt{3} M_P$, which is about one order of magnitude
larger than $(\alpha/ 4 \pi) F / M_{GUT}$. Nevertheless, some
scenarios like the sequestering mechanism discussed in
Ref.~\cite{Dermisek:2006qj} lead to a suppression of the
supergravity contributions, such that the MSSM soft terms
predominantly arise from messenger loops. 
In this section we shall assume that such a mechanism is
at work. The gaugino and scalar soft masses then receive
two types of contributions: {\it (i)} gauge messenger contributions arising
from loops of $X$ and $Y$ gauge fields, and {\it (ii)} chiral messenger
contributions from {\it (a)} the weak triplet and colour octet components
of $\Sigma$, and {\it (b)} the colour Higgs triplet and antitriplet:
\bea
  m^2_\chi (M_{GUT}) & = & \Delta_{GM} m^2_\chi
    + \Delta_{(\Sigma_3, \Sigma_8)} m^2_\chi + \Delta_{(T, \bar T)} m^2_\chi\, ,  \\
  M_a (M_{GUT}) & = & \Delta_{GM} M_a
    + \Delta_{(\Sigma_3, \Sigma_8)} M_a + \Delta_{(T, \bar T)} M_a\, .
\eea
Each contribution can be computed using the wave function
renormalization technique~\cite{Kaplunovsky:1983yx,Giudice:1997ni}.
For gauge messengers we find, in agreement with the literature:
\bea
  \Delta_{GM} m^2_\chi (M_{GUT}) & = & \left( \frac{\alpha_{GUT}}{4\pi} \right)^2
    2 \left[ b_{SU(5)} \Delta c^\chi - \sum_a \Delta b_a c^\chi_a \right]
    \left| \frac{F}{M_{GUT}} \right|^2 ,  \hskip .5cm  \\
  \Delta_{GM} M_a (M_{GUT}) & = & -\, \frac{\alpha_{GUT}}{4\pi}\, \Delta b_a\,
    \frac{F}{M_{GUT}}\, ,
\eea
where $\Delta b_a = 2 (C[SU(5)] - C[G_a])$ is the contribution of
the massive $X$ and $Y$ vector multiplets to the beta function
coefficient of the gauge group $G_a$ ($G_a = SU(3)_c, SU(2)_L, U(1)_Y$
for $a=1,2,3$, respectively),
$\Delta c^\chi = c^\chi_{SU(5)} - \sum_a c^\chi_a$ is the difference between
the second Casimir coefficients of the $SU(5)$ and \SM\, representations
to which $\chi$ belongs, the normalization of the hypercharge generator
is the SU(5) one and we have identified $v = M_{GUT}$.
For $\Sigma_3$, $\Sigma_8$ and $(T, \bar T)$, one obtains the standard 
chiral messenger formula, with\footnote{It should be stressed that
$F/M |_{\Sigma_3}$ and $F/M |_{\Sigma_8}$ are model dependent.
The factor of $2$ relative to $F/M_{GUT}$ is specific to the quartic
superpotential of Eq.~(\ref{eq:W_Sigma}).}
$F/M |_{\Sigma_3} = F/M |_{\Sigma_8} = 2 F/M_{GUT}$.
Note that the triplet contribution depends on how the doublet-triplet
splitting is realized; here we assume that this is done
with a single fine-tuning in the way described in Section~\ref{dt},
in which case one also has $F/M |_{(T, \bar T)} = 2 F/M_{GUT}$.

The resulting gaugino and scalar soft masses at the messenger scale
are given in the table below, together with the values of each individual
contribution, in units of
\beq
  m_{GM}\, =\, \frac{\alpha_{GUT}}{4\pi}\, \frac{F}{M_{GUT}}\, .
\eeq

$$
\begin{array}{|c|c|c|c|c|c|c|c|c|}
\hline
 m^2_\chi / m^2_{GM} & Q & U^c & E^c & L & D^c & H_u, H_d  \\
\hline
 \Delta_{GM} & -11 &  -4 & 6 & -3 & -6 & -3  \\
\hline
 \Delta_{(\Sigma_3, \Sigma_8)} & 44 &  32 & 0 & 12 & 32 & 12  \\
\hline
 \Delta_{(T, \bar T)} & \frac{804}{75} &  \frac{864}{75} & \frac{48}{25} & \frac{12}{25}
 & \frac{816}{75} & \frac{12}{25}  \\
\hline
 {\rm total} & \frac{1093}{25} &  \frac{988}{25} & \frac{198}{25} & \frac{237}{25}
 & \frac{922}{25} & \frac{237}{25}  \\
\hline
\end{array}
\hskip .5cm
\begin{array}{|c|c|c|c|c|c|}
\hline
 M_a / m_{GM} & M_3 & M_2 & M_1  \\
\hline
 \Delta_{GM} & -4 &  -6 & -10  \\
\hline
 \Delta_{(\Sigma_3, \Sigma_8)} & 6 &  4 & 0  \\
\hline
 \Delta_{(T, \bar T)} & 2 &  0 & \frac{4}{5}  \\
\hline
 {\rm total} & 4 & -2 & - \frac{46}{5}  \\
\hline
\end{array}
$$

\noi As for A-terms, they only receive contributions from gauge messengers:
\beq
  A_\chi (M_{GUT})\, =\, \frac{\alpha_{GUT}}{4\pi}\,
    2 \Delta c^\chi\, \frac{F}{M_{GUT}}\, ,
\eeq
with $A_u = A_Q + A_{U^c} + A_{H_u}$, etc, yielding:
\beq
  A_u (M_{GUT}) = 10\, m_{GM}\, , \
  A_d (M_{GUT}) = 8\, m_{GM}\, , \
  A_e (M_{GUT}) = 12\, m_{GM}\, .
\eeq
We have assumed the minimal field content, with only the adjoint
$\Sigma$ and a pair of Higgs fields in the $5 \oplus \bar 5$ representation
(giving $b_{SU(5)} = 3$),
and neglected the running of the gauge couplings between the
various messenger scales $M_{GUT}$, $m_3$, $m_8$ and $m_T$.

A few comments are in order. First, the negative gauge messenger
contributions to scalar squared masses are smaller than the
positive contributions from chiral messengers, so that there is no
tachyonic mass at the messenger scale\footnote{This is different from
Ref.~\cite{Dermisek:2006qj}, in which chiral messenger contributions
are omitted, and the renormalization group running is crucial to
make all sfermion squared masses positive at the electroweak scale.}.
Second,
the soft terms are rather large in units of $m_{GM}$,
implying that even a moderate suppression of supergravity corrections
may be enough for gauge-mediated contributions to dominate.

Let us briefly comment on the physical superpartner spectrum
obtained by running the above boundary conditions at the messenger
scale down to the electroweak scale.
If we require squarks to be lighter than $2 \TeV$ or so, the lightest
Higgs boson mass will be typically below $120 \GeV$, despite the
fact that a rather large $A_t$ is generated at the messenger scale.
A Higgs mass around the value indicated by the
ATLAS~\cite{Collaboration:2012si} and CMS~\cite{Collaboration:2012tx}
experiments requires significantly heavier squarks and sleptons
(unless other contributions to the top squark A-term are invoked,
e.g. from a non-renormalizable superpotential term
$10^3_F 10^3_F\, \Sigma\, 5_H / M_P$). As an example, we display
in Table~\ref{tab:spectrum} the spectrum corresponding
to the choice $m_{GM} = 960 \GeV$, $\tan \beta = 25$ and $\mu > 0$
(with $\mu$ and $B\mu$ determined from radiative electroweak
symmetry breaking) obtained with the help of the code
SuSpect~\cite{Djouadi:2002ze}. 
The parameters $m_3 = 4m_8$ and $m_T$ were chosen in
such a way that gauge coupling unification occurs at
$M_{GUT} = 10^{17} \GeV$, but the effect of the corresponding
thresholds on the running of the soft terms has been neglected.
A noticeable feature of this spectrum is that the LSP is the wino
(the gravitino mass being $6.4 \TeV$), with a mass in the right
ballpark to account for the observed dark matter relic density.
Unfortunately, this spectrum cannot be tested at the LHC.

\begin{table}[t]
\begin{tabular}{|c|c|c|c|c|c|c|c|c|c|c|}
\hline
  h & A & $H^0$ & $H^\pm$ & $\tilde \chi^\pm_1$ & $\tilde \chi^\pm_2$
  & $\tilde \chi^0_1$ & $\tilde \chi^0_2$ & $\tilde \chi^0_3$
  & $\tilde \chi^0_4$ & $\tilde g$  \\
\hline
  124.75  &    6267.8  &    6267.8  &    6268.2
  & 1750 &     6002 &   1750 &     3743 &       6001 &     6002 & 8296  \\
\hline
\end{tabular} \par
\hskip .6cm \begin{tabular}{|c|c|c|c|c|c|c|c|}
\hline
  $\tilde t_1 $ & $\tilde t_2$ & $\tilde u_1, \tilde c_1$ & $\tilde u_2, \tilde c_2$
  & $\tilde b_1$ & $\tilde b_2$ & $\tilde d_1, \tilde s_1$ & $\tilde d_2, \tilde s_2$  \\
\hline
  7421 &   8350 &   9275 &  9295  & 8330 & 8726 &  8949 & 9296  \\
\hline
\end{tabular} \par
\hskip .6cm \begin{tabular}{|c|c|c|c|c|c|}
\hline
  $\tilde \tau_1$ & $\tilde \tau_2$ & $\tilde e_1, \tilde \mu_1$ & $\tilde e_2, \tilde \mu_2$
  & $\tilde \nu_\tau$ & $\tilde \nu_e, \tilde \nu_\mu$  \\
\hline
  3050  &  3361  &   3557 &  4236  & 3053 &    3556  \\
\hline
\end{tabular}
\caption{Superpartner masses (in GeV) predicted by the SU(5) model
of Section~\ref{su5}, assuming suppressed supergravity contributions
and doublet-triplet splitting through a single fine-tuning.
The input parameters are $M_{GUT} = 10^{17} \GeV$, $b_{SU(5)} = 3$,
$m_{GM} = 960 \GeV$, $\tan \beta = 25$ and $\mbox{sign}(\mu) = +$.}
\label{tab:spectrum}
\end{table}
%

%%%%%%%%%%%%%%%%%%%%%
\section{\label{conclusions}Conclusions}
%%%%%%%%%%%%%%%%%%%%%

One of the main obstructions in constructing predictive supersymmetric
Grand Unified Theories is our ignorance of the way supersymmetry is broken. 
This leaves proton decay rates as well as the fermion mass
fitting undetermined, since these observables depend on the way 
supersymmetry is broken and mediated to the MSSM sector.
A crucial step towards predictive supersymmetric unification is therefore
to implement supersymmetry breaking and its mediation in realistic GUTs.

In this paper, we showed that it is possible to break simultaneously
supersymmetry and the gauge symmetry with a single field, in spite
of the absence of a tree-level flat direction. The local supersymmetry-breaking
minimum is induced radiatively in the effective potential, far away
from the gauge- and supersymmetry-preserving global minimum, in a
region where the tree-level potential is almost flat. The condition for this
mechanism to work is a fine-tuned correlation between the superpotential
parameters. Supersymmetry breaking is then transmitted to the MSSM
through both gauge and chiral messenger loops, thus avoiding the
suppression of gaugino masses characteristic of direct gauge mediation
models~\cite{Komargodski:2009jf}. The use of a single field ensures
that no dangerous tachyonic scalar masses are generated at the one-loop level.

We studied the implementation of this mechanism in an SU(5)
model with a single adjoint representation. A generic prediction
is that intermediate states are unavoidable, which raises the GUT
scale and allows for a larger Higgs triplet mass, as preferred
by the experimental lower bound on the proton lifetime.
In the particular case of a quartic superpotential, the Higgs triplet
mass is even predicted, but a strong fine-tuning among the couplings
of the quartic terms is needed to achieve unification below the Planck
scale. An unfortunate consequence of the increase of the GUT scale
is that supergravity contributions to the MSSM soft terms will generally
dominate over the gauge-mediated ones, unless one relies on
a mechanism like no-scale supergravity or conformal sequestering.
Only in this case can one obtain definite predictions for the superpartner
spectrum. These were presented in Section~\ref{sec:spectrum}.
We also discussed model-building aspects related to the quark-lepton
mass relations and to the doublet-triplet splitting, which present
significant differences with the case of separate supersymmetry
and gauge symmetry breaking sectors.

Similar predictions to the ones found in this paper are obtained in a
supergravity model with a single SU(5) adjoint, in which supersymmetry
and the gauge symmetry are broken at the tree level~\cite{Bajc:2006pa}.
It would be interesting to investigate how the mechanism that we have
proposed would be affected by local supersymmetry.
Also, it remains to be studied how it can be embedded in a higher-rank
Grand Unified Theory.

%%%%%%%%%%%%%%%%%%%%%%%%%%%%%%%%%
\section*{Acknowledgments}
We thank K.S. Babu for discussions.
This work has been supported in part by the Slovenian Research
Agency, by the Agence Nationale de la Recherche under Grant
ANR 2010 BLANC 0413 01, by the European Commission under
Contract PITN-GA-2009-237920, and by the French-Slovenian
programme Proteus under Contracts  BI-FR/10-11-PROTEUS-14
and 22773RG. BB and SL acknowledge partial financial support from
the NSF under Grant No. 1066293 and the hospitality of the Aspen
Center for Physics. SL thanks the Galileo Galilei Institute for Theoretical
Physics for hospitality and the INFN for partial support.
%%%%%%%%%%%%%%%%%%%%%%%%%%%%%%%%%

%%%%%%%%%%%

\end{document}